\pdfoutput=1
\documentclass[fleqn,usenatbib]{mnras}

\usepackage{newtxtext,newtxmath}
\usepackage[T1]{fontenc}

\usepackage{amsfonts}
\usepackage{amsmath}
\usepackage{mathrsfs}
\usepackage{epsfig}
\usepackage{graphicx}
\usepackage{url}
\usepackage[utf8]{inputenc} 
\usepackage{graphicx}
\usepackage{epsf}
\usepackage{comment}
\usepackage{aas_macros}
\usepackage{slashed}
\usepackage[caption=false]{subfig}

\title{Have hierarchical three-body mergers been detected by LIGO/Virgo?}

\author[D. Veske et al.]{
Do\u{g}a Veske,$^{1}$\thanks{E-mail: dv2397@columbia.edu}
Zsuzsa M\'arka,$^{2}$
Andrew G. Sullivan,$^{1}$
Imre Bartos,$^{3}$
\newauthor
K. Rainer Corley,$^{1,2}$
Johan Samsing$^{4}$
and Szabolcs M\'arka$^{1}$
\\
$^{1}$Department of Physics, Columbia University in the City of New York, 550 W 120th St., New York, NY 10027, USA\\
$^{2}$Columbia Astrophysics Laboratory, Columbia University in the City of New York, 550 W 120th St., New York, NY 10027, USA\\
$^{3}$Department of Physics, University of Florida, PO Box 118440, Gainesville, FL 32611-8440, USA\\
$^{4}$Niels Bohr International Academy, The Niels Bohr Institute, Blegdamsvej 17, DK-2100, Copenhagen, Denmark
}

\begin{document}
\label{firstpage}
\pagerange{\pageref{firstpage}--\pageref{lastpage}}
\maketitle

\begin{abstract}
One of the proposed channels of binary black hole mergers involves dynamical interactions of three black holes. In such scenarios, it is possible
that all three black holes merge in a so-called hierarchical merger chain, where two of the black holes merge first and then their remnant subsequently merges
with the remaining single black hole. Depending on the dynamical environment, it is possible that both mergers will appear within the observable time window.
Here we perform a search for such merger pairs in the public available LIGO and Virgo data from the O1/O2 runs. Using a frequentist p-value assignment statistics we do not find any significant merger pair candidates, the most significant being GW170809-GW151012 pair. Assuming no observed candidates in O3/O4, we derive upper
limits on merger pairs to be $\sim11-110\ {\rm year^{-1}Gpc^{-3}}$, corresponding to a rate that relative to the total merger rate is $\sim 0.1-1.0$.
From this we argue that both a detection and a non-detection within the next few years can be used to put useful constraints on some dynamical progenitor models.
\end{abstract}


\section{Introduction}
\label{sec:Introduction}

The LIGO Scientific Collaboration and the Virgo Collaboration
have publicly announced properties of 10 binary black hole (BBH) mergers from the first and second observing runs (O1 and O2) in the gravitational wave (GW) catalog GWTC-1 \citep{Abbott_2019}. Individual groups have also performed searches on the open data from O1 and O2 and found additional merger candidates \citep{venumadhav2019new,zackay2019detecting,Nitz_2019,Zackay_2019}. From those, \cite{venumadhav2019new,zackay2019detecting,Zackay_2019} report 8 more BBH mergers, total of 18 BBH mergers, whose samples are publicly available at \url{https://github.com/jroulet/O2\_samples} (IAS-Princeton mergers hereafter).
The set of confirmed events have been used to constrain e.g. general relativity and its possible modifications \citep[e.g.][]{2019PhRvD.100j4036A}; however, how and where the BBHs form in our Universe are still major unsolved questions. There are several plausible formation scenarios, including
field binaries \citep{2012ApJ...759...52D, 2013ApJ...779...72D, 2015ApJ...806..263D, 2016ApJ...819..108B,
2016Natur.534..512B, 2017ApJ...836...39S, 2017ApJ...845..173M, 2018ApJ...863....7R, 2018ApJ...862L...3S},
chemically homogeneous binary evolution \citep{10.1093/mnras/stw1219,10.1093/mnras/stw379,refId0},
dense stellar clusters \citep{2000ApJ...528L..17P,
2010MNRAS.402..371B, 2013MNRAS.435.1358T, 2014MNRAS.440.2714B,
2015PhRvL.115e1101R, 2016PhRvD..93h4029R, 2016ApJ...824L...8R,
2017MNRAS.464L..36A, 2017MNRAS.469.4665P}, 
active galactic nuclei (AGN) discs \citep{2017ApJ...835..165B,  2017MNRAS.464..946S, 2018ApJ...866...66M,2019ApJ...876..122Y},
galactic nuclei (GN) \citep{2009MNRAS.395.2127O, 2015MNRAS.448..754H,
2016ApJ...828...77V, 2016ApJ...831..187A, 2016MNRAS.460.3494S, 2017arXiv170609896H, 2018ApJ...865....2H},
very massive stellar mergers \citep{Loeb:2016, Woosley:2016, Janiuk+2017, DOrazioLoeb:2017},
and single-single GW captures of primordial black holes \citep{2016PhRvL.116t1301B, 2016PhRvD..94h4013C,
2016PhRvL.117f1101S, 2016PhRvD..94h3504C}.
The question is; how do we observationally distinguish these merger channels from each other? Recent work has shown that the measured BH spin \citep{2016ApJ...832L...2R}, mass spectrum \citep{2017ApJ...846...82Z,2019PhRvL.123r1101Y}, and orbital eccentricity \citep{2014ApJ...784...71S, 2017ApJ...840L..14S, 2018ApJ...853..140S, 2018PhRvD..97j3014S, 2018ApJ...855..124S, 2018MNRAS.tmp.2223S, 2019ApJ...871...91Z, 2018PhRvD..98l3005R, 2019arXiv190711231S, 2019PhRvD.100d3010S} can be used. In addition, indirect probes of BH populations have also been suggested; for example, stellar tidal disruption events can shed light on the BBH orbital distribution and corresponding merger rate in dense clusters \citep[e.g.][]{2019PhRvD.100d3009S}, or spatial correlations with host galaxies \citep{2017NatCo...8..831B}.

In this paper we perform the first search for a feature we denote `hierarchical merger chains' that are unique to highly
dynamical environments \citep[e.g.][]{2018MNRAS.476.1548S, 10.1093/mnras/sty2249}.
The most likely scenario of a hierarchical merger chain is the interaction of three BHs, $\{ BH_{1}, BH_{2}, BH_{3}\}$, that undergo two subsequent mergers; the first between $\{ BH_{1}, BH_{2}\}$ and the second between $\{BH_{12},BH_{3}\}$, where $BH_{12}$ is the BH formed in the first merger.
Such hierarchical merger chains have been shown to form in e.g. globular clusters (GCs) as a result of binary-single interactions. In this case, the first merger happens during the three-body interaction when the BHs are still bound to each other, which makes it possible for the merger remnant to subsequently merge with the remaining single BH \citep{2018MNRAS.476.1548S,10.1093/mnras/sty2249}. Fig. \ref{fig:DMfig} illustrates schematically this scenario.
Such few-body interactions are not restricted to GCs, but can also happen in e.g. AGN discs \citep[e.g.][]{2019arXiv191208218T}. 
Interestingly, under certain orbital configurations, both the first and the second merger can show up as detectable GW signals within the observational time window \citep[e.g.][]{10.1093/mnras/sty2249}. The hierarchical merger chain scenario can therefore be observationally constrained, and can as a result be used to directly probe the dynamics leading to the assembly of GW sources.

With this motivation, we here look for hierarchical merger pair events in the public O1 and O2 data from LIGO and Virgo. For this, we present a new algorithm to identify merger pairs, the simplest example of a hierarchical merger chain, and use it to search for such events in the public GWTC-1 catalogue and in the IAS-Princeton sample.

The paper is organized as follows. In Section \ref{sec:search} we describe our search method, and in Section \ref{sec:results} we present the corresponding results. Finally, we conclude our work in Section \ref{sec:conclusion}.

\section{Search}
\label{sec:search}

\begin{figure}
    \centering
    \includegraphics[width=0.9\columnwidth]{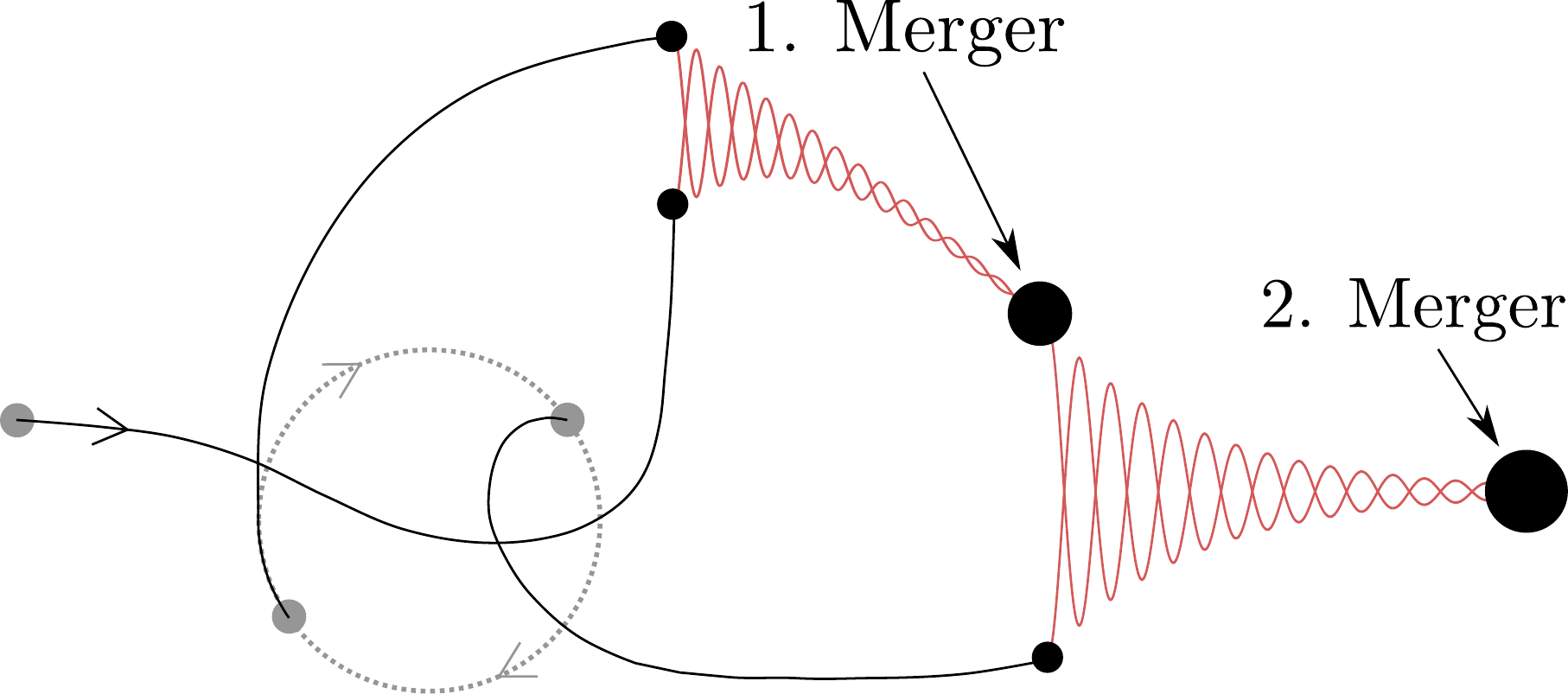}
    \caption{Illustration of a hierarchical merger chain, where two subsequent BBH mergers form from a single three-body interaction. The interaction progresses from left to right, where the BH tracks are highlighted with black thin lines. As seen, the initial configuration is a binary interacting with an incoming single (grey dots). During the interaction, two of three BHs merge, after which the product merges with the remaining single \citep{10.1093/mnras/sty2249}. In this paper we search for such BBH merger pairs.}
    \label{fig:DMfig}
\end{figure}

In this section we describe our methods for searching for GW merger pairs originating from three-body interactions like the one
shown in Fig. \ref{fig:DMfig}. 

\subsection{Parameters}

Our search is based on a frequentist p-value assignment by using a test statistic (TS). As Neyman-Pearson's lemma suggests \citep{doi:10.1098/rsta.1933.0009}, we choose our TS to be the ratio of the likelihood of the signal hypothesis to the likelihood of the null hypothesis; where we define our null hypothesis \(H_0\) as having two unrelated mergers, and our signal hypothesis \(H_s\) as having two related mergers originating from a three-body interaction. We use 3 parameters of the BBH mergers for calculating the likelihood ratio:
\begin{itemize}
    \item {\it Mass estimates:} One of the initial BH masses in the second merger
    should agree with the final mass of the BH formed in the first merger.
    \item {\it Correct time order:} The first merger, as defined by the mass difference, should happen before the second merger.
    \item {\it Localization:} Both the first and the second merger must
    originate from the same spatial location.
\end{itemize}
Using these three parameters our TS is
\begin{equation}
TS=\begin{cases}\frac{\mathcal{L}(M_f,m_{1,s},m_{2,s},V_f,V_s|H_s)}{\mathcal{L}(M_f,m_{1,s},m_{2,s},V_f,V_s|H_0)} &, t_f<t_s\\
0 &, t_f\geq t_s
\end{cases}
\end{equation}
where \(\mathcal{L}\) represents the likelihoods of the parameters for each hypothesis, $M$ represents the final mass estimate, $m_1$ and $m_2$ represent the mass estimates of the merging BHs, $V$ represents the spatial localization, and $t$ represents the merger times. Subscripts $f$ and $s$ represent the first and second merger, respectively. We do not use the spins of the BHs due to
large uncertainties in the spin measurements \cite[e.g.][]{Abbott_2019}; however,
we do hope this becomes possible later, as spin adds an additional strong constraint (the BH formed in the first merger typically appears in the second merger with a spin of $\sim 0.7$ \citep[e.g.][]{2007PhRvD..76f4034B,2017ApJ...840L..24F}).

For writing down the likelihoods we assume that the individual BH masses in the first merger follow a power law distribution with index -2.35 between 5-50\(M_{\odot}\) (denoted as $\mathscr{M}_i$) \citep{PhysRevX.6.041015}. We further assume 5\% of the total initial BH mass is radiated during merger, as suggested by previous detections and theory \cite[e.g.][]{Abbott_2019}. Hence, for BHs which are a result of a previous merger
the corresponding mass spectrum is the self-convolution of the $\mathscr{M}_i$ mass spectrum (denoted as $\mathscr{M}_c$) with its values reduced by 5\%. We marginalize over these mass distributions and a $r^2$ distribution for distance ($r$) when calculating the likelihoods. We are well aware that different dynamical channels predict different BH mass distributions; however, we do find that our results do not strongly depend on the chosen model. The power of the search mainly comes from comparing two detections with each other rather than comparing them to a prior distribution.The full expression for the likelihood ratio is given in the Appendix.

\subsection{Generating the background distribution}
\label{bg}
Our significance test is based on a frequentist p-value assignment via comparison with a background distribution. In order to have the background distribution, we perform BBH merger simulations and localize them with BAYESTAR \citep{PhysRevD.93.024013,Singer_2016}. The simulations assume that the mass of BHs that are not a result of a previous merger is drawn independently from our assumed initial BH mass distribution \(\mathscr{M}_i\). The mergers are distributed uniformly in comoving volume, and the orientation of their orbital axes are uniformly randomized. We assume the BH spins to be aligned with the orbital axis and we don't include precession \citep{2019MNRAS.488.4459C}. We use the reduced-order-model (ROM) SEOBNRv4 waveforms \citep{PhysRevD.95.044028}, and the cosmological parameters from the nine-year WMAP observations \citep{2013ApJS..208...19H}. The simulated detection pairs are made at O2 sensitivity for different detector combinations corresponding to first and second merger detected by either the LIGO Hanford-LIGO Livingston (HL) combination or the LIGO Hanford-LIGO Livingston-Virgo (HLV) combination. We denote the pairs that are both detected by HL as HL-HL, both by HLV as HLV-HLV, first by HL and second by HLV as HL-HLV, and first by HLV and second by HL as HLV-HL.

In order to construct the background distributions for the likelihood ratios, we need the same inputs as real detections.
For this, we first assume that there is 5\% mass loss in the merger to have a central value for the final mass. Second, in order to include realistic detection uncertainties, we broaden the exact masses to triangular distributions whose variances depend on the signal-to-noise ratio (SNR) of the detections and the distributions' modes are the exact masses. We use the triangular distributions for imitating the asymmetry of the estimates in the real detections around the median \citep{Abbott_2019}. For determining the upper and lower bounds of the triangular likelihood distributions of masses we use a linear fit whose parameters are obtained by fitting a line to the relative 90\% confidence intervals of the mass estimate likelihoods of real detections (which is obtained by dividing the posterior distribution to prior distribution from the parameter estimation samples) as a function of detection SNR. This fit is done separately for both component masses and the final masses. The minimum relative uncertainty is bounded at 5\% which is the lowest uncertainty from real detections \citep{Abbott_2019}.

Before moving on the results of our search, in order to estimate the possible capability of our search, we created artificial triple merger pairs by drawing the initial BH masses from the \(\mathscr{M}_i\) spectrum and distributing the pairs uniformly in comoving volume. For the best case scenario, HLV-HLV detection, we found that \(\sim 90\%\) of the merger pairs have more than one sided \(3\sigma\) (p-value\(\leq1/740\)) significance. For the HLV-HL, HL-HLV and HL-HL scenarios, the ratios of the pairs that have more than \(3\sigma\) significance to the total number of pairs are \(\sim 70\%,\ 60\%\) and \(20\%\) respectively. This shows the importance of having better localization with the \(3^{\rm rd}\) detector for this analysis.
\section{Results and Discussion}
\label{sec:results}
In this section we show and discuss our results for the 18 BBH mergers. We use both the samples for the 10 GWTC-1 and 18 IAS-Princeton mergers and find p-values for each separately. These merger counts give us a total of 45 possible hierarchical merger pair combinations for GWTC-1 and 153 for the IAS-Princeton sample.

\begin{figure}
    \centering
    \includegraphics[width=\columnwidth]{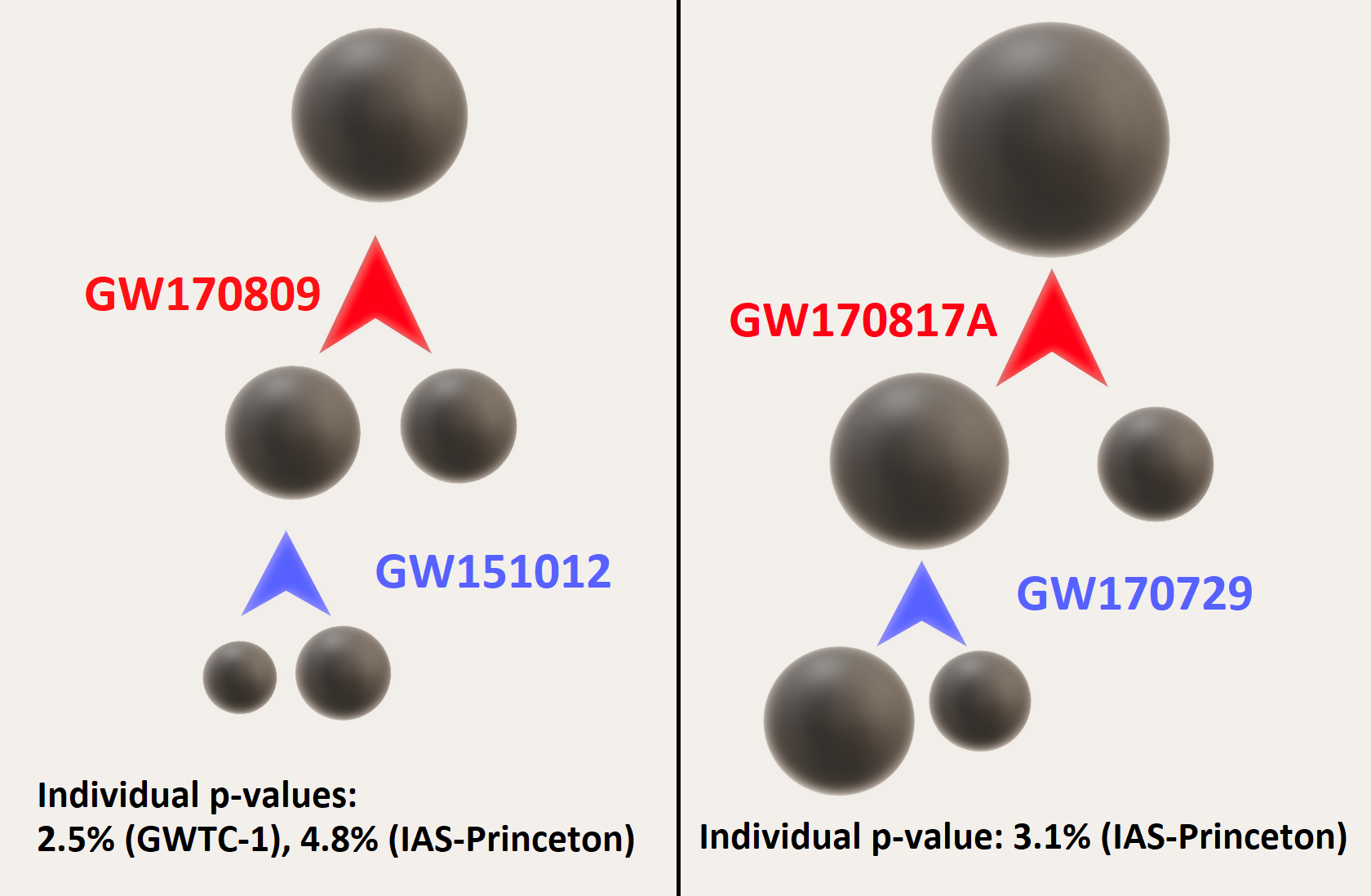}
    \caption{The consecutive merger scenarios for the two most significant event pairs with their individual p-values for IAS-Princeton and GWTC-1 samples. The second pair is only present in the IAS sample since GW170817A is not listed in GWTC-1.}
    \label{fig:triples}
\end{figure}

\subsection{Event Pair Significance}
\label{sec:Event Pair Significance}
In Fig. \ref{fig:triples} we show the 2 most significant event pairs from our search. The most significant merger pair GW151012 (first merger) and GW170809 (second merger) has an individual p-value of 2.5\% from the GWTC-1 sample, meaning that only 2.5\% of the background event pairs are more significant than this. Its p-value from the IAS-Princeton sample is 4.8\%, slightly higher. The significance of the pair comes from the matching of the primary mass of GW170809 with the mass of the final black hole of GW151012. However the primary mass of GW170809 ($\sim 35M_{\odot}$) is well below the (hypothesized) pair-instability mass limit and GW170809 was not thought of a potential hierarchical merger result.

Our second most significant event involves GW170729 (first merger) and GW170817A (second merger), with individual p-value of 3.1\%. GW170817A's primary mass exceeds the (hypothesized) pair-instability mass limit (its median is $\sim 50M_{\odot}$ with support up to $\sim80M_{\odot}$) suggesting it could be the result of a previous merger \citep{Gayathri_2020}. Our analysis suggests that GW170729 is a plausible previous merger for GW170817A in the hierarchical merger scenario, through the GW1710817A's primary black hole, as in the GW170809-GW151012 pair. However, as explained at the end of the section, after one accounts for the multiple hypothesis testing correction, none of the event pairs analysed can be considered significant enough for a decisive discovery.

GW170729, itself also has a primary mass estimation similar to GW170817A's primary mass which indicates it may also be result of a previous merger \citep{Abbott_2019,2019PhRvL.123r1101Y}\citep[cf.][]{2020RNAAS...4....2K}. However, the significance of event pairs involving GW170729 as the second merger in our analysis are lower; the two most significant pairs being GW170729-GW151012 and GW170729-GW170403. The individual p-values are 5.5\% (GWTC-1) and 17\% (IAS-Princeton) for GW170729-GW151012, and 11\% (IAS-Princeton) for GW170729-GW170403.

Finally, we notify that as the number of events increases, we will inevitably have low p-value event pairs. To account for this, one has to include a `multiple hypothesis correction', which in our case brings a factor of 198 (the number of analysed merger pairs) to the individual p-values. After this correction, none of the event pairs can be considered significant. When we compare the significance of GW170809-GW151012 pair with our artificially generated triple pairs detected with HL-HLV combination, we find that \(\sim 98\%\) of artificially generated pairs to be more significant than the GW170809-GW151012 pair. Similarly for the GW170817A-GW170729 pair, \(\sim 99\%\) of the artificially generated HLV-HLV pairs are more significant.

\subsection{Limits on hierarchical triple merger rates}
\label{sec:Limits on hierarchical triple merger rates}

We start by estimating the upper limits on the rate density of hierarchical merger pairs given the absence of an observed pair during O1 and O2. For this we assume that the first mergers in the hierarchical chain scenario are Poisson point processes with a uniform rate density per comoving volume, \(R\), and that the temporal difference between the two mergers, $t_{12}$, follows a power law distribution \(P(t_{12}<T) \propto (T/t_{max})^\alpha\), where $t_{\max}$ ($T \leq t_{\max}$) and $\alpha$ ($\alpha > 0$) are parameters that are linked to the underlying dynamical process \citep[e.g.][]{10.1093/mnras/sty2249}. We further assume the duty cycle of each given time period is the same during the observing runs,
i.e., we do not consider the non-uniformity of running times during the runs. The duty cycle for having at least two operating detectors during O1 is 42.8\% and during O2 is 46.4\% \citep{Vallisneri_2015,Abbott_2019}. Studies have shown that about half of all BBH mergers forming during three-body interactions will appear with an eccentricity $e>0.1$ at 10 Hz \citep{2019arXiv190711231S,Rodriguez_2018}. However, current matched filter search template banks only include circular orbits \citep{LIGOeccentric} (except a recent study on binary neutron star mergers \citep{alex2019search}). Non-template based searches are able to recover eccentric binaries \citep{2019PhRvD.100f4064A}, but with somewhat lower sensitivity compared to that of template based searches for circular binaries for the masses considered here. Hence, for simplicity, we consider a 50\% loss of efficiency as well. Together with this loss, we denote the overall duty cycles as \(\kappa_1\) and \(\kappa_2\), respectively for O1 and O2, and the O1 duration by \(\Delta t_1\), the O2 duration by \(\Delta t_2\), and the time in between O1 and O2 by \(\Delta t_0\) (O1 lasted about 4 months, O2 lasted about 9 months and they had about 10 months in between). The search comoving volumes are denoted for O1 and O2 by $C_1$ and $C_2$, respectively. These two volumes, $C_1$ and $C_2$, we estimate by (i) using the ratios of the ranges of the LIGO instruments in the O1, O2 and O3 runs; (ii) the search comoving volume for the O3 run in \cite{collaboration2013prospects}; (iii) neglecting the contribution to the search comoving volume in O3 by Virgo (due to having less than the half range of LIGO detectors), and (iv) assuming independent 70\% duty cycles for the LIGO detectors in O3 \citep{collaboration2013prospects}.
We estimate $C_1$ to be 0.07 ${\rm Gpc^3year/year}$ and $C_2$ to be 0.14 ${\rm Gpc^3year/year}$. Following this model
we then calculate the probability \(\mathcal{P}\) of not seeing a hierarchical merger pair during O1 and O2 (The full expression for \(\mathcal{P}\) is found in the Appendix).
Results are presented in Fig. \ref{fig:rate}, which shows the frequentist 90\% upper limit for the rate density \(R\) that satisfies \(\mathcal{P}=0.1\), for different values of $t_{max}$ and $\alpha$. We have chosen $t_{max}$ values between $10$ and $10^{7}$ years which are the expected order magnitudes for prompt mergers and non-prompt mergers (see \cite{10.1093/mnras/sty2249}). Hence, those represent the limiting cases of all mergers being prompt and non-prompt. As seen, the upper rate density varies between \(\sim 150-210\ {\rm year^{-1}Gpc^{-3}}\) for our chosen range of values.

We now investigate the expected future limits for triple hierarchical mergers assuming a null result when the third observing run of LIGO and Virgo (O3), and planned fourth observing run (O4) with KAGRA \citep{PhysRevD.88.043007}, also are included in our search. O3 started on April 1st, 2019, and is planned to have 12 months of observing duration, with a one month break in October 2019. Although O4 dates remain fluid, it is estimated to be in between 2021/2022-2022/2023 \citep{collaboration2013prospects}. For our study we assume O3 and O4 to last for a year, with O4 starting in January 2022. The comoving search volumes in O3 and O4 are estimated to be 0.34 ${\rm Gpc^3year/year}$ and 1.5 ${\rm Gpc^3year/year}$, respectively. Although it will be more accurate to include the contribution from Virgo to these volumes, we here neglect its contribution to the duty cycles in a conservative manner and assume 70\% independent duty cycles for the LIGO detectors \citep{collaboration2013prospects}. We adopt the median expected BBH merger detection counts from \cite{collaboration2013prospects}, which are 17 and 79 for O3 and O4 respectively. Our derived lowest limits with the inclusion of O3 and O4 is shown in Fig. \ref{fig:rate}. As seen, the rate densities are now \(\sim 11-110\ {\rm year^{-1}Gpc^{-3}}\). 

We end our analysis by investigating the upper limits for the fractional contribution from the first mergers of the hierarchical triple mergers to the total BBH merger rate. For the detection number and duration of the O1 and O2 runs, then at 90\% confidence, the upper limits of the fractional contribution for the model parameters we consider in Fig. \ref{fig:rate} are all $\approx 1$. We get more informative upper limits when we consider absence of merger pairs in the O3 and O4 runs as illustrated in the lower panel of Fig. \ref{fig:rate}. As seen, the upper limits now vary between \(\sim 0.1-1\).

\begin{figure}
    \centering
    {\includegraphics[width=\columnwidth]{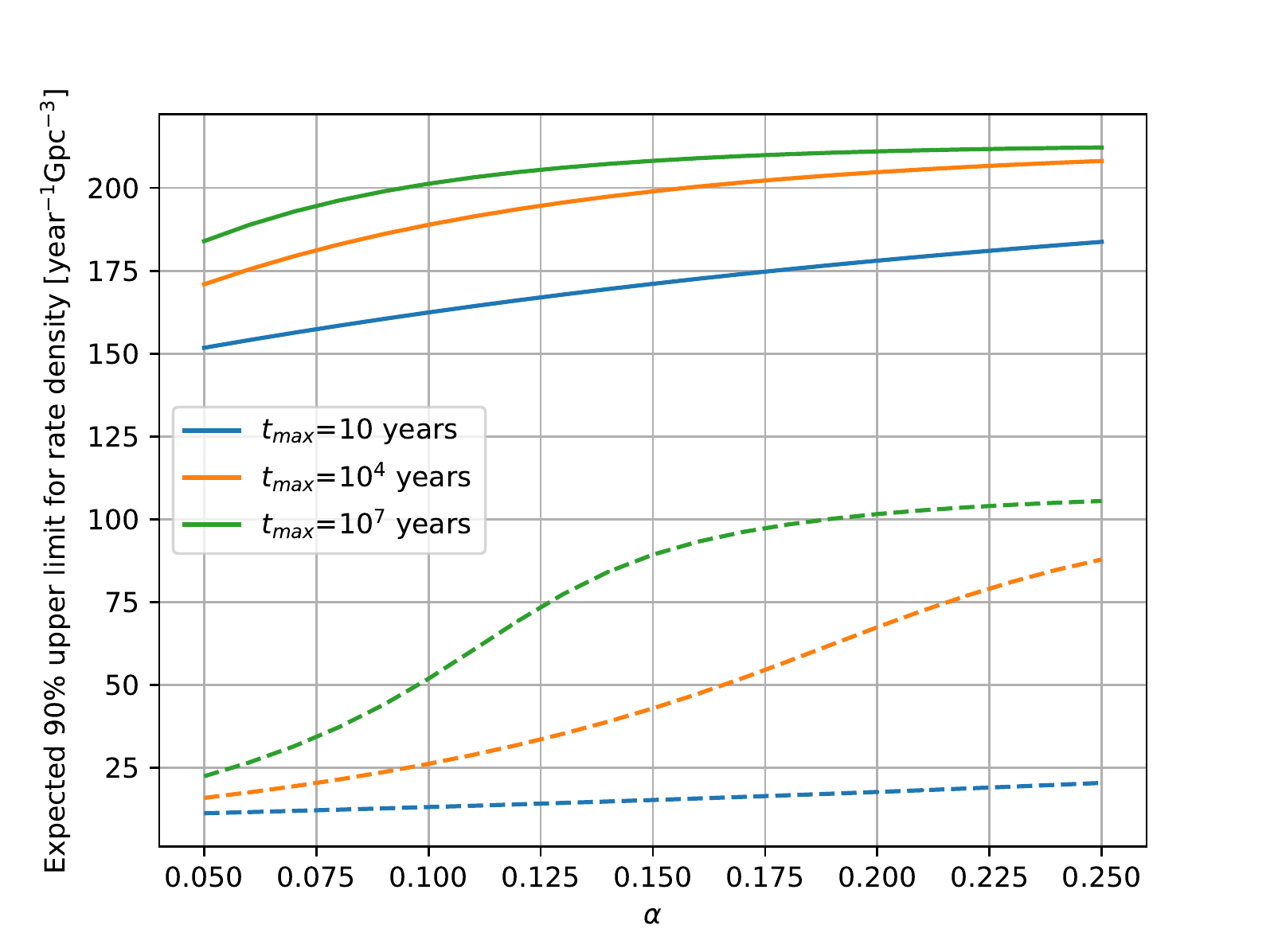}}
    {\includegraphics[width=\columnwidth]{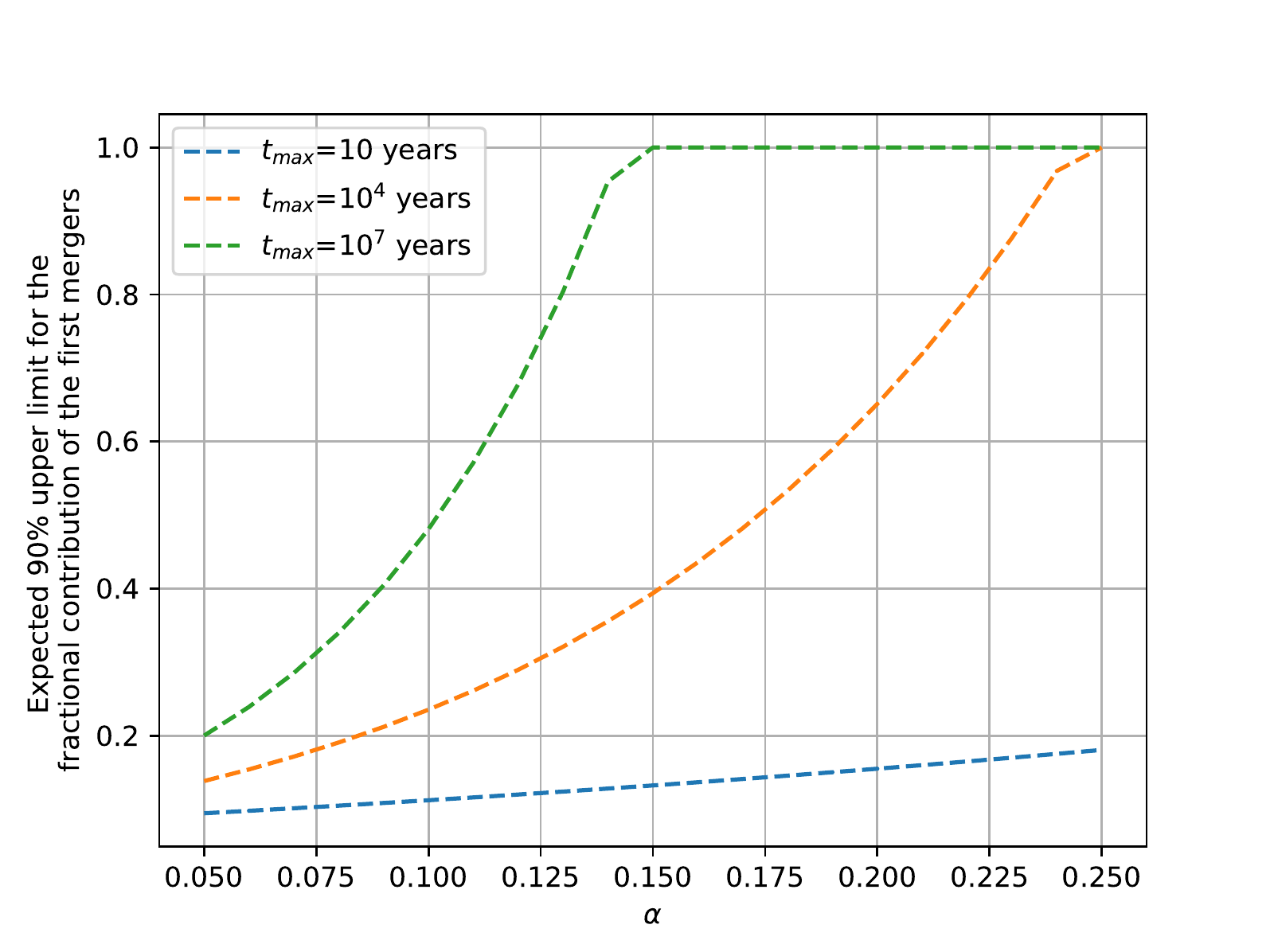}}
    \caption{Expected 90\% upper limit of density (\emph{top}) and fractional contribution to the total observed BBH merger rate (\emph{bottom}) for the first mergers of the triple hierarchical mergers. \emph{Solid} lines show the rate densities considering the absence of a significant event pair in O1 and O2 runs. \emph{Dashed} lines show the rate densities assuming the absence of a significant event pair when O3 and O4 are also included.}
    \label{fig:rate}
\end{figure}

Finally, we stress that our rate estimates from this section are associated with large uncertainties, mainly due to unknowns in the underlying dynamical model.
For example, the functional shape of our adopted $P(t_{12} < T)$-model from Section \ref{sec:Limits on hierarchical triple merger rates}, depends in general on both the BH mass hierarchy, the exact underlying dynamics, the initial mass function, as well as on the individual spins of the BHs \citep[e.g.][]{10.1093/mnras/sty2249}; all of which are unknown components. Another aspect is how the rate limit depends on other measurable parameters, such as orbital eccentricity and BH spin. For example, in \cite{10.1093/mnras/sty2249} it was argued that most hierarchical three-body merger chains are associated with high eccentricity; a search for eccentric BBH mergers, as the one performed in \cite{2019MNRAS.490.5210R}, can therefore be used to put tight constraints on this scenario. Another example, is the effective spin parameter, $\chi_{\rm eff}$, which was used to argue that the primary BH of GW170729 is likely not a result of a previous BBH merger despite its relative high mass and spin \citep{2020RNAAS...4....2K}.
However, we are actively working on improving our search algorithm both through the inclusion of eccentricity and spin. Having a fast and accurate pipeline searching for correlated events might also be useful for putting constraints on gravitationally lensed events.

\section{Conclusion}
\label{sec:conclusion}
We presented a search method (Section \ref{sec:search}) for detecting hierarchical GW merger pair events resulting from binary-single interactions (see Fig. \ref{fig:DMfig}), and applied it to the public available O1/O2 data from the LIGO and Virgo collaborations.
Using a frequentist p-value assignment statistics we do not find any significant
GW merger candidates in the data that originate from a hierarchical binary-single merger chain (Section \ref{sec:Event Pair Significance}).
Using a simple model for describing the time between the first and second merger (Section \ref{sec:Limits on hierarchical triple merger rates}),
we estimated the upper limit on the rate of hierarchical mergers from binary-single interactions
from the O1/O2 runs to be \(\sim150-210\ {\rm year^{-1}Gpc^{-3}}\) for varying parameter values of our time-difference model. Assuming
no significant merger pairs in the O3/O4 runs we find the upper limit reduces to \(\sim11-110\ {\rm year^{-1}Gpc^{-3}}\), corresponding to a rate
that relative to the total merger rate is $\sim 0.1-1.0$. The theoretical predicted rate of hierarchical GW merger pair events is highly uncertain;
however, we have argued and shown that both a detection and a non-detection of merger pairs can provide useful
constraints on the origin of BBH mergers. In future work we plan on including both eccentricity and BH spin parameters in our search for
hierarchical GW merger pair events. Moreover, considering the expectancy of such events happening in dense environments, known AGNs or other plausibly related dense environments can also be used to correlate with the spatial reconstruction of the events in the search.

\section*{Acknowledgments}
The authors are grateful for the useful feedback of Christopher Berry and Jolien Creighton.
The authors thank the University of Florida and Columbia University in the City of New York for their generous support.
The Columbia Experimental Gravity group is grateful for the generous support of the National Science Foundation under grant PHY-1708028. D.V. is grateful to the Ph.D. grant of the Fulbright foreign student program. A.S. is grateful for the support of the Columbia College Science Research Fellows program.
J.S. acknowledges support from the European Unions Horizon 2020 research and
innovation programme under the Marie Sklodowska-Curie grant agreement No. 844629. The authors are grateful to Leo Singer of GSFC for the BAYESTAR package and his valuable help with our use case scenario. This research has made use of data, software and/or web tools obtained from the Gravitational Wave Open Science Center (\url{https://www.gw-openscience.org}), a service of LIGO Laboratory, the LIGO Scientific Collaboration and the Virgo Collaboration. LIGO is funded by the U.S. National Science Foundation. Virgo is funded by the French Centre National de Recherche Scientifique (CNRS), the Italian Istituto Nazionale della Fisica Nucleare (INFN) and the Dutch Nikhef, with contributions by Polish and Hungarian institutes.
\section*{Data Availability}
The data underlying this article were accessed from \url{http://dx.doi.org/10.7935/KSX7-QQ51} and \url{https://github.com/jroulet/O2_samples}. The derived data generated in this research will be shared on reasonable request to the corresponding author.
\bibliographystyle{mnras}
\bibliography{Refs}

\label{lastpage}

\onecolumn
\appendix

\section{Likelihood ratio}
All BBH mergers are assumed to be uniformly distributed in comoving volume. In this case the likelihood ratio becomes
\begin{multline}
\frac{\mathcal{L}(M_f,m_{1,s},m_{2,s},V_f,V_s|H_s)}{\mathcal{L}(M_f,m_{1,s},m_{2,s},V_f,V_s|H_0)}=\frac{\int P(M_f,m_{1,s},m_{2,s}|m',H_s)P(m'|H_s)dm' \int P(V_f,V_s|r,H_s)P(r|H_s)drd\mathbf{\Omega}}{\int P(M_f,m_{1,s},m_{2,s}|H_0)dm' \int P(V_f,V_s|H_0)drd\mathbf{\Omega}} \\
=\frac{ \int \frac{P(r|V_f)P(r|V_s)}{r^2}drd\mathbf{\Omega} \sum_{\substack{x,y=1,2 \\ x\neq y}}\int \frac{P(m'|M_f)}{P_f(m')}\frac{P(m'|m_{x,s})}{P_{x,s}(m')}\mathscr{M}_c(m')dm'\int \frac{P(m'|m_{y,s})}{P_{y,s}(m')}\mathscr{M}_i(m')dm'}{\int \frac{P(m'|M_f)}{P_f(m')}\mathscr{M}_c(m')dm'\int \frac{P(m'|m_{1,s})}{P_{1,s}(m')}\mathscr{M}_i(m')dm'\int \frac{P(m'|m_{2,s})}{P_{2,s}(m')}\mathscr{M}_i(m')dm'}
\end{multline}
where $m'$, $r$ and $\mathbf{\Omega}$ are the integration variables for mass, distance and sky location. \(P_f(m')\), \(P_{1,s}(m')\) and \(P_{2,s}(m')\) are the mass priors used in the parameter estimation. We take these from the parameter estimation sample released in GWTC-1 \citep{parameter_estimation} and from \url{https://github.com/jroulet/O2\_samples} for the IAS-Princeton sample. The integrals over the spatial localization in the denominator equals unity and are therefore not written. The summed terms in the numerator represent either of the BHs in the second merger resulting from the first merger.

\section{Probability $\mathcal{P}$}

Here we write the probability \(\mathcal{P}\) of not seeing a hierarchical merger pair for the parameters \(R\), \(t_{max}\), \(\alpha\), \(\kappa_1\), \(\kappa_2\), \(\Delta t_1\), \(\Delta t_2\), \( \Delta t_0\), and with the number of seen events, $n_i$, during O1 ($n_1 = 3$) and O2 ($n_2 = 7$). The condition of not seeing a pair of hierarchical mergers is to see at most one of the mergers in the pair.
\begin{multline}
\mathcal{P}=\bigg[\sum_{i=0}^{n_1} Poisson(i,\kappa_1 R\Delta t_1 C_1) \frac{i!}{{\Delta t_1}^i}  \int_0^{\Delta t_1}\int_{0}^{\tau_i}...\int_{0}^{\tau_2}[1-\kappa_2(\frac{\Delta t_1+\Delta t_2+\Delta t_0-\tau_1}{t_{max}})^\alpha+\kappa_2(\frac{\Delta t_1+\Delta t_0-\tau_1}{t_{max}})^\alpha-\kappa_1(\frac{\Delta t_1-\tau_1}{t_{max}})^\alpha]\times \\... \times [1-\kappa_2(\frac{\Delta t_1+\Delta t_2+\Delta t_0-\tau_{i-1}}{t_{max}})^\alpha+\kappa_2(\frac{\Delta t_1+\Delta t_0-\tau_{i-1}}{t_{max}})^\alpha-\kappa_1(\frac{\Delta t_1-\tau_{i-1}}{t_{max}})^\alpha] \\ \times [1-\kappa_2(\frac{\Delta t_1+\Delta t_2+\Delta t_0-\tau_i}{t_{max}})^\alpha+\kappa_2(\frac{\Delta t_1+\Delta t_0-\tau_i}{t_{max}})^\alpha-\kappa_1(\frac{\Delta t_1-\tau_i}{t_{max}})^\alpha]d\tau_1...d\tau_{i-1}d\tau_i\bigg] \\ 
\times \bigg[\sum_{i=0}^{n_2} Poisson(i,\kappa_2 R\Delta t_2 C_2) \frac{i!}{{\Delta t_2}^i}  \int_0^{\Delta t_2}\int_{0}^{\tau_i}...\int_{0}^{\tau_2}[1-\kappa_2(\frac{\Delta t_2-\tau_1}{t_{max}})^\alpha] \times ... \times [1-\kappa_2(\frac{\Delta t_2-\tau_{i-1}}{t_{max}})^\alpha]\times [1-\kappa_2(\frac{\Delta t_2-\tau_i}{t_{max}})^\alpha]d\tau_1...d\tau_{i-1}d\tau_i\bigg]
\label{eq:long}
\end{multline}
with \(Poisson(n,k)\) being the probability of seeing \(n\) events from the Poisson point process with mean \(k\). \(\frac{i!}{\Delta t}\) is the value of joint probability distribution of Poisson arrival times given that there are \(i\) events in time interval \(\Delta t\). The integrals give the probability of not seeing any of the second mergers of \(i\) observed first mergers during the observation times. The first term in Eq. \eqref{eq:long} gives the probability of not seeing an hierarchical merger pair whose first event can happen during O1 and second event can happen during O1 or O2. The second term gives the probability of not seeing an hierarchical merger pair whose both mergers can happen during O2. Multiplication of them gives us the probability of not seeing an hierarchical pair during O1 and O2. We use the integral identity
\begin{equation}
\int_0^a\int_0^{\tau_i}...\int_0^{\tau_2}f(\tau_1)\times ... \times f(\tau_{i-1})\times f(\tau_i)d\tau_1...d\tau_{i-1}d\tau_i = (\int_0^a f(\tau_1)d\tau_1)^i\frac{1}{i!}
\end{equation}
to simplify the expression for \(\mathcal{P}\).
\begin{multline}
\mathcal{P}=\bigg[\sum_{i=0}^{n_1} Poisson(i,\kappa_1 R\Delta t_1 C_1) \frac{1}{{\Delta t_1}^i} \big[\int_0^{\Delta t_1} [1-\kappa_2(\frac{\Delta t_1+\Delta t_2+\Delta t_0-\tau_1}{t_{max}})^\alpha+\kappa_2(\frac{\Delta t_1+\Delta t_0-\tau_1}{t_{max}})^\alpha-\kappa_1(\frac{\Delta t_1-\tau_1}{t_{max}})^\alpha] d\tau_1\big]^i\bigg] \\ \times \bigg[\sum_{i=0}^{n_2} Poisson(i,\kappa_2 R\Delta t_2 C_2) \frac{1}{{\Delta t_2}^i}\big[\int_0^{\Delta t_2}[1-\kappa_2(\frac{\Delta t_2-\tau_1}{t_{max}})^\alpha]d\tau_1\big]^i\bigg]
\label{eq:short}
\end{multline}

\end{document}